\documentstyle[12pt]{article}
\newcommand{\be}{\begin{equation}}
\newcommand{\ee}{\end{equation}}
\newcommand{\bea}{\begin{eqnarray}}
\newcommand{\eea}{\end{eqnarray}}
\newcommand{\n}{\bar{n}}

\begin{document}
 \begin{titlepage}

\begin{flushright}
CERN-TH.6889/93
\end{flushright}
\vspace{20 mm}

\begin{center}
{\huge Quantum mechanics, common sense}
\vspace{1 mm}
{\huge and}
\vspace{1mm}
{\huge the black hole information paradox}
\end{center}

\vspace{10 mm}

\begin{center}
Ulf H. Danielsson and Marcelo Schiffer\footnote{Supported by a John
 Stewart Bell Fellowship}\\
Theory Division, CERN, CH-1211 Geneva 23, Switzerland.

\end{center}

\vspace{20 mm}

\begin{center}
{\large Abstract}
\end{center}

The purpose of this paper is to analyse, in the light of information
theory and with the arsenal of (elementary) quantum mechanics (EPR
correlations, copying machines, teleportation, mixing produced in
sub-systems owing to a trace operation, etc.) the scenarios
available  on the market to resolve the so-called black-hole information
paradox. We shall conclude that the only   plausible ones      are those
where either the unitary evolution of quantum mechanics is given up, in
which information leaks continuously in the course of black-hole
evaporation through non-local processes, or those in which the world is
polluted by an infinite number of  meta-stable remnants.

\vspace{6cm}
\begin{flushleft}
CERN-TH.6889/93 \\
May 1993
\end{flushleft}
 \end{titlepage}
\newpage

\section{Introduction}

In this paper we will discuss the black-hole information paradox,
first discovered by Hawking \cite{hawk}. As discussed by him,
when  a pure state has collapsed to form a black hole, it will
later evolve into a mixed one as the outcome of the complete evaporation
of the black hole. In the wake of this observation, a fierce controversy
emerged in the literature. 't Hooft \cite{hooft}
proposed, as a way out of the paradox,
that some unknown mechanism could provide the needed correlation
between incoming and outgoing radiation to save the     unitary
evolution of quantum states.             Nevertheless,
as became   increasingly clear during the past year or so,
a resolution of the paradox will need a much better understanding
of the interplay between gravity and quantum mechanics than is
currently at hand. In this context,
a lot has been learnt from   studies of
two-dimensional black holes initiated by Callan et al. in
\cite{CGSH}.
It might even be           that the information paradox
is our best clue to the elusive quantum gravity theory.
It is therefore of extreme importance to have a thorough understanding
of this paradox, as free of model-dependent technicalities as possible.

With the arsenary of elementary quantum mechanics and some
 information theory we will illustrate the paradox.
Our simple analysis will shed  some light on the very
nature of the paradox and define the properties that any solution
must possess. In particular, we will consider the point of view that
a black hole is a ``quantum object", somehow implying that our
usual intuition of what is wrong and right in physics is not
applicable. This typically suggests that EPR-like correlations
are important and that this would circumvene the standard
arguments leading to the paradox.
We will find that no such easy way out is possible.

We will begin by analysing the problem using elementary quantum
mechanics. Then,
in section 3, we will use information theory to
derive some simple results concerning the way in which information
may be stored in black holes.

In section 4 we will arrive at a standard set of possibilities, however
with a much better understanding of why none    of these can be
conservative in the sense of not involving fundamentally new
phenomena.

\section{Quantum Copy Rights}

In this section we will consider limitations on the possible resolutions
of the information paradox due to quantum mechanics. It is
important to see why certain obvious suggestions do not work.

We will consider a situation where the information is ``copied"
before the in-falling matter crosses the horizon. In this way
it is made
available to the Hawking radiation. In fact, for an outside
observer, the in-falling matter will not be seen to cross
the horizon until very late. For an eternal black hole it
would never be seen to cross. Hence one might think that
all information is conveniently stored and accessible.
Still, the act of making a copy is necessary if
the original is assumed to
continue through the horizon and into the black hole. This, in
turn, is based
on our expectation, due to the equivalence principle, that the
horizon does not have any exceptional local properties capable
of completely reflecting all information. If this had been the case
we would have had a very simple resolution of the paradox at hand.

In general,
both the original and the copy may experience
a unitary transformation through some scattering matrices.
The process is schematically
\be
| \psi \rangle  \rightarrow |\psi _{B} \rangle   \otimes |\psi _{O}
 \rangle \, .
 \label{prod}
\ee
Since the final state is a direct product between the internal black-
hole state $| \psi _{B} \rangle$ and the outside state
$|\psi _{O} \rangle$,
there are no correlations  between
the inside and the outside. Hence, if we ignore the inside, i.e. take the
trace, no mixing will result on the outside. There would then be no
loss of
information. Is this a possible scenario?
Unfortunately (\ref{prod}) is forbidden. One cannot copy quantum
states in this way \cite{wozu}.
The proof goes as follows. Let us assume the state to be copied to be
a spin 1/2 particle with states $|\downarrow\rangle $ and
$|\uparrow\rangle $.
For simplicity we will ignore the state of the copying machine
itself. This can be taken into account, \cite{wozu}, with no change
in the conclusions.
The copying process must
be described by some unitary operator $U$.
Let us assume that the copying process works for states that  are
purely up or down.
By linearity we then have
\be
U\left[  (a|\downarrow\rangle  +b|\uparrow\rangle  \right]
= a  |\downarrow\rangle |\downarrow
\rangle
+b |\uparrow\rangle |\uparrow\rangle  .
\ee
However, the desired state
\be
(a|\downarrow\rangle  +
b|\uparrow\rangle )(a|\downarrow\rangle +b|\uparrow\rangle ) =
a^{2}|\downarrow\rangle |\downarrow \rangle
+b^{2}|\uparrow\rangle |\uparrow\rangle
+ab(|\downarrow\rangle |\uparrow\rangle  +
|\uparrow\rangle |\downarrow\rangle )
\ee
cannot result for general $a$ and $b$, since $U$ produces no states
$|\downarrow \rangle |\uparrow \rangle $
or $|\uparrow \rangle |\downarrow \rangle $.
We conclude that even if one can construct
a $U$ which works for a given state, the same $U$ will not work
for all states.
In a sense, $U$ is too good at making copies!
The correlations are always perfect in the up/down basis.
Hence taking the trace
over one subsystem produces a {\it maximal} mixing in the other
subsystem and
hence a loss of information. In fact, in this case all information is
stored in the correlations.

Now, can this perfect correlation be exploited? If we, given the outside
 state,
always know the inside through these perfect correlations, clearly there
 can
not be any loss of information. It would be silly to take the trace over
the inside, since it is {\it identical} to the outside, and
interpret    this as true entropy.
The situation recalls of
the EPR-phenomenon. Is this the way to solve the paradox?
Again the suggestion does not work. The reason is that
the correlation cannot be perfect for all states in all bases. This is
clearly needed if we are allowed to make any measurement that
we want. Consideronce more our example:
\be
U(a|\downarrow \rangle +b|\uparrow \rangle ) = a|\downarrow \rangle
 |\downarrow \rangle +
b|\uparrow \rangle |\uparrow \rangle .
\ee
Use
\be
|\downarrow \rangle
=\frac{1}{\sqrt{2}} (|\rightarrow \rangle -|\leftarrow \rangle )
\;\;\;\;\;   ,
|\uparrow \rangle
=\frac{1}{\sqrt{2}} (|\rightarrow \rangle +|\leftarrow \rangle )
\ee
to get
$$
a|\downarrow \rangle |\downarrow \rangle  +
b|\uparrow \rangle |\uparrow \rangle
$$
\be
=\frac{a+b}{2} (
|\rightarrow \rangle |\rightarrow \rangle  +
|\leftarrow \rangle |\leftarrow \rangle ) +\frac{a-b}{2}(
|\rightarrow \rangle |\leftarrow \rangle  +
|\leftarrow \rangle |\rightarrow \rangle )  .
\ee
It is only when ($a=-b$) $a=b$ that the (anti-) correlation is perfect.
 In the EPR case this means that it is only for singlet states that the
anti-correlation is perfect in all bases.
Hence, since the correlation is not perfect in general,
we are forced to take the trace. At any rate, for a given unknown
 state, an EPR- related state cannot be obtained through a
unitary copying process that works for a general state.

For completeness we should note one loophole in the above argument. This
is the case of black-hole hair as discussed in \cite{har}.
According to these
ideas there are an infinite number of conserved quantities in the world
whoose conservation protects unitarity. For this to be the case,
{\it everything} needs to be conserved, which amounts to say that the
 world
is an integrable system. This means that there are superselection rules
that  forbid superpositions. Compare the superselection rule for
electric charge.
In the presence of these superselection rules the above
argument will not hold.
On the other hand, one faces the difficult problem of reconstructing
    quantum mechanics as we know it, starting with this barren
universe.

It seems, therefore, that we have to cope with the fact that
information does
cross the horizon and is at least temporarily hidden from the outside
observer.
The questions then are: if, when and how is the information restored?
In the next section we will consider the possibility that the information
is stored not locally, in the black hole,
but rather in its correlations with the environment.

\section{How to Store Information}

As is well known, there is a fundamental objection
from QFT to the idea that
the information is stored in a black-hole remnant. Low-mass objects with
a huge number of internal states would suffer from enormous production
 rates
completely inconsistent with observations. This argument is not
qualitatively changed if we take into account that the remnants may
slowly evaporate and disappear. Since very little energy is available
and a lot of information must be transmitted, the needed time is very
long and the remnant effectively stable as far as the
argument is concerned  \cite{presk}.

There have
been attempts to construct remnants that would not  have this defect
\cite{banks}. However, these attempts seem to run into inevitable problems
\cite{gidd}. We will not consider this further.

In an interesting paper  \cite{wil}, it has been suggested that the
 information
need not be stored locally in the remnant, which implies the above
 problem, but
rather in its correlations with the outside world. This would then, it
seems, point at a conservative resolution of the paradox. It is important
to note that the correlations we have in mind are correlations between
the emitted radiation and the black hole, not correlations between
radiation emitted at different times. The reason that the latter
is not so relevant  is that until the late-time radiation is emitted,
the information still has to reside somewhere. This must be {\it inside}
the black hole. This is because, as we      proved in the previous
section, given some reasonable assumptions,
there will always be information crossing the
horizon that is impossible for the Hawking radiation to copy.
As we will see, and comment on           later on, the correlations
can be restored to the Hawking radiation (e.g. between radiation
emitted at different times) only through non-local processes.

Below, we will
analyse  the situation using information theory. We will consider two
coupled systems $1$ and $2$ with basis $|n\rangle _{1}$, $n=1, ...,
 N_{1}$, and
$|m\rangle _{2}$, $m=1, ..., N_{2}$, where
 $N_{2} \geq N_{1}$. We will assume that
the initial state of the combined system is pure, i.e.that
\be
| \psi \rangle =
\sum _{n,m=1}^{N_{1},N_{2}} A_{nm} |n\rangle _{1}|m\rangle _{2}  \, .
\ee
The corresponding pure density matrix is
\be
\rho = \sum _{n,m,k,l}^{N_{1},N_{2}} A_{nm} |n\rangle _{1}
|m\rangle _{21} \langle p|_{2} \langle q|A_{pq}^{*}   \, .
\ee
{}From this one may construct reduced, in general mixed,
 density matrices for the
individual subsystems $1$ and $2$. For $1$ we obtain
\be
\rho _{1} =\sum _{j,n,p}^{N_{1}, N_{2}}A_{nj}A_{pj}^{*}
|n\rangle _{11} \langle p|
\ee
and for $2$ we get
\be
\rho _{2} =\sum _{j,m,q}^{N_{1},N_{2}} A_{jm}A_{jq}^{*}
|m\rangle _{22} \langle q|  \, .
\ee

Information will be defined as follows \cite{zur}
\be
I=I_{max} +Tr\rho \log \rho
\ee
where $S=-Tr\rho \log \rho$ and $I_{max}=S_{max}$.
The entropy, $S$, is to be thought of as a lack of information.
Note that $S=0
\Rightarrow I=I_{max}$ and $S=S_{max} \Rightarrow I=0$.
If the number of states is
$N$, we have $S_{max} =-N\frac{1}{N} \log \frac{1}{N} = \log N$,
where $\rho =\frac{1}{N}$ for all states. So,
\be
I=\log N +Tr\rho \log \rho    .
\ee
With two subsystems we have
$$
I_{1} =\log N_{1} +Tr \rho _{1} \log \rho _{1}  \, ,
$$
$$
I_{2} =\log N_{2} +Tr \rho _{2} \log \rho _{2}   \, ,
$$
\be
I_{tot} =\log N_{1} N_{2}+Tr \rho  \log \rho    \, ,
\label{inf}
\ee
and
\be
I_{tot} =I_{1}+I_{2}+I_{12}   \, ,
\ee
which {\it defines} $I_{12}$, the information content of the
 correlations.

With a pure total state the total information is maximized (i.e.
the entropy is zero)
\be
I_{tot} = \log N_{1} +\log N_{2}  .
\ee
What then can be said about the information content of the separate
 systems
$1$ and $2$? Clearly $I_{1,max}=\log N_{1}$ and $I_{2,max}=\log N_{2}$,
but what else can we know? Below we will prove that
\be
I_{2,min} = \log N_{2} - \log N_{1} .   \label{imin}
\ee
The proof is simple: $A_{nm}$ is an $N_{1} \times N_{2}$ matrix ($N_{1}$
rows and $N_{2}$ columns); $\rho _{1} = AA^{\dagger}$ is an
$N_{1} \times
N_{1}$ matrix and $\rho _{2} = (A^{\dagger}A)^{*}$
an $N_{2} \times N_{2}$.
We first prove that $A^{\dagger}A$ has at least $N_{2}-N_{1}$ zero
 eigenvalues.
To do so, let us
construct the $N_{2}\times N_{2}$ matrix $\tilde{A}$ by adding
$N_{2}-N_{1}$ rows of zeros. Clearly $A^{\dagger}A = \tilde{A}^{\dagger}
\tilde{A}$ and  $\tilde{A}$ has then
at least $N_{2}-N_{1}$ zero eigenvalues by construction. If
$\tilde{A}$ is diagonalized, so is
$\tilde{A}^{\dagger}\tilde{A}$. Therefore we find
that $\tilde{A}^{\dagger}\tilde{A}$,
and also $A^{\dagger}A$, have at least $N_{2}-
N_{1}$ zero eigenvalues. To minimize $I_{2}$ we must put $\rho _{2} =
\frac{1}{N_{1}}$ for the remaining $N_{2}-(N_{2}-N_{1})=N_{1}$ non-zero
eigenvalues. Then (\ref{imin}) follows.

The result (\ref{imin}) is very reasonable. A little tracing in a small
subsystem cannot produce a lot of entropy, or loss of information,
in the rest of the system.

Let us now pretend that system $2$ is the outside
world, containing the Hawking
radiation, and that system $1$ is the interior of the black hole.
If we find that there is very little information in $2$, i.e. $I_{2}
 \sim 0$,
we must conclude that $N_{1} \sim N_{2}$.
That is, the number of internal states must be very large. It might,
however, still be the case that the information is not stored in system
$1$ but in the correlations, i.e. $I_{1}=0$ and $I_{12} \neq 0$. The
important point is that if the information is to be stored in the
correlations between the subsystems, each of the subsystems must still
 have
the {\it capacity} to store (half of)
the information. This must be the case even
if the capacity is not used!

Let us now be more precise and relate the above reasoning to a more
realistic model of a black hole. When the black hole is formed,
we assume that the total system is in a pure state. There is
information stored in the outside world, the black hole itself, and
necessarily also in correlations. The latter is a consequence of
the non-existence of perfect copying machines, as we saw in
the previous section. As the black hole begins to evaporate,
entropy will be produced in the outside world subsystem. Our
objective is to estimate a lower limit on this entropy if
we ignore back reaction or any other transfer of
information to the Hawking radiation.
The total entropy  carried by the radiation per unit time during the
evaporation is then
\begin{equation}
\dot{S}= \sum_j \int \frac{d\omega}{2\pi}S_j(\omega)  \, ,
\label{entropy}
\end{equation}
where $d\omega /2\pi$ is the number of phase cells per unit time that
emanate from the black hole                               and $S_j$
is the entropy in a given field mode of the j-th species \cite{lan}
\begin{equation}
S_j(\omega)=-\left[ {\n}_j \ln {\n}_j \mp
(1 \pm {\n}_j)\ln (1 \pm {\n}_j) \right] .
\end{equation}
Here and in what follows, the lower and upper signs apply for fermions
and bosons, respectively. On the other hand, the mean number of quanta
emitted in a given mode by the back hole is \cite{hawk}:
\begin{equation}
\n=\frac{\Gamma}{e^x \mp 1} \, ,
\label{spontaneous}
\end{equation}
 with $x=\hbar \omega/T_{\rm bh}$ and  $\Gamma$ is the black
hole absorptivity.

The calculation of the entropy flux in eq.
(\ref{entropy}) by means of the above equations has to be carried
out numerically, because the black-hole absorption coefficient cannot
be cast in a closed form. Here, we  borrow  Page's \cite{page,page'}
result where he calculated $\dot{S}$ numerically for a mixture of
three species of neutrinos and antineutrinos, photons and gravitons
\begin{equation}
\dot{S}= 1.619 \frac{\dot{E}}{T_{\mbox{\tiny bh}}}.
\end{equation}
Integrating this equation, we obtain the amount of mixing in the
radiation produced along  the black-hole history. Together with
eqs. (13 and 16) we can write the relations
\begin{equation}
\ln N_{1} >
S_{\mbox{\tiny radiation}}= 1.619 S_{\mbox{\tiny bh}}.
\end{equation}
So,
the presence of entropy in the outside world puts a lower limit
on the number of necessary states of the black hole.
Note that this really is a lower limit: there is also entropy
initially, before the evaporation has begun, which is
due to the always present correlations between
what went in and what stayed behind. This may generally be of the
same order.

These relations teach us two things.
First, if the information has {\it not} been returned through
Hawking radiation as the black hole approaches the Planck mass, then the
remnant has to have an enormous number of internal states to save
unitarity. The information might be stored in correlations, as in
\cite{wil}, but this does not solve the remnant problem.
Secondly, if we
decide  to follow the rules of quantum mechanics, we have to
seriously interpret $e^{S_{\mbox{\tiny bh}}}$ as the number of black-hole
 quantum states. The black hole must make
full use of its quantum states in order for  the information that it
subtracted from the enviroment to    be momentarily stored  either in
these states themselves or in correlations.
Furthermore, we have learned from the previous discussion  the
 information in
question cannot wait until the last moments of evaporation to be
restored. Accordingly, it has to leak steadily in the course of
 black-hole evaporation.

A popular
point of view is that back reaction could transfer the information
from the in-falling matter forming the black hole to the Hawking
radiation. As we have seen, there are
two sources of entropy for the outside world.
One is the matter that  formed the black hole, the other one
is the Hawking radiation, or rather the negative energy part
that  falls into the black-hole. The idea of back reaction
suggests that the Hawking
pair production is influenced, in such a way that
the two potential sources of entropy
conspire so that at the end
no entropy is produced.
As we have seen  in the previous section,
such a process can {\it never} be perfect, if, as is commonly
assumed, it is possible to travel into a black hole without losing
one's memory.
In this connection, it has recently been shown that
stimulated emission (bosons) and the exclusion principle (fermions)
are two such mechanisms,     providing
an imperfect correlation between incoming
and outgoing radiations, which allows a partial transfer of the
information content of the former to the latter \cite{marc}.
Hence, the only remaining possibility is non-local information transfer.

\section{Three Possibilities}

In  view of the previous discussion,
we see only three possible solutions to the paradox.

{\bf I}. Give up unitary quantum mechanics.

{\bf II}. Find a way to get along with the remnants. No such possibility
seems to exist at the moment  \cite{gidd}.

{\bf III}. The information is restored as the black hole evaporates. This
requires non-local effects.

We will discuss the third possibility in a little more detail.
The non-locality
which is needed is not just the standard non-locality of
quantum mechanics.
This would have been in the spirit of correlations, and we have just
shown that this is not enough. Instead, one needs a rue information
flow from behind the (apparent) horizon.

It is amusing to compare this situation with the idea of Bennet {\it et
al} \cite{tele} on
teleportation. There a state is destroyed at one point in space time
only to reappear at another. Two kinds of information transfer
are needed: one nonlocal EPR-like piece and one classical piece,
which must respect the causal structure of space-time. More precisely,
the sender and the receiver are each equipped with the members of
EPR pairs. The sender brings its EPR particles together with
the state to be teleportated. He then makes some measurements on
the combined system. The results are then sent to the receiver
who, with this knowledge, may reconstruct the teleportated state.
This is
also the case here. In fact, the parallel is rather complete.
The EPR pairs are the pair-produced Hawking radiation, with one
particle escaping and the other one venturing into the black hole.
The problem is that the second part of the information transfer,
which
is crucial as we have seen, is troubled by the horizon.
Now,
the relevant horizon is an apparent horizon, which means that
escape is possible but has to be delayed  until very late. At this
later stage the storage capacity of the black hole has
necessarily decreased, unless we contemplate alternative {\bf II}.
Therefore the information must either be destroyed, alternative
{\bf I}, or transferred from the interior and the correlations to
the exterior, alternative {\bf III}.

In the latter case, the question is how?
 If we trust    the correspondence
 principle, no spectacular quantum gravity effects could occur  in the
 outgoing radiation when the black hole is large with respect to
 Planckian scales. So  it seems that the black-hole must make use of
 nonlocal effects through its quantum states for transferring the
 information in question.

It has been recently suggested \cite{muk,bel}, based on information
theoretic premisses, that the black-hole event horizon is quantized
in units of Planck length squared and, furthermore, similarly to
what happens in atomic physics, the leakage of information is made
possible by        transitions among various quantum black-hole states
(black hole-spectroscopy) \cite{bel}. Let us analyse, from the
information theoretic point of view, whether this mechanism could account
for the information flow needed to solve the paradox. That is to say,
whether the entropy associated with
the different transitions from a given
state to the ground state (total evaporation) is comparable with the
information the black hole has subtracted from the environment. In order to
estimate this, let us assume that the black hole is in an eigenstate
of event horizon area $|A,x\rangle$, where $x$ stands for
the set of quantum numbers accounting for the corresponding
degeneracy $e^{\frac{A}{4}}$ for a given $A$.
Now, the transition probability from level $|A,x\rangle$ to
$|A',x'\rangle $, for {\em any} $x$ and $x'$, must be proportional to
the ratio between the degeneracy of the levels in question.
Accordingly, the probability of transition of going from level $A$ to
$A'$ cannot strongly depend on whether the transition occurs directly or
if it proceeds through intermediate states. The reason is that
in order to estimate the transition probability from the initial to the
final state in the case of cascading, we have to multiply all the
intermediate transition probabilities, assuming that these
are statistically independent. After multiplying all these probabilities
and cancelling out the intermediate degeneracies, we end up with the
ratio between the degeneracies of the final and initial states, exactly
as if the transition had occurred in one step. Thus, in order to obtain
 an estimation of the information that could be transferred
to the environment by means of
the black-hole spectral lines, should they exist, we assume that all
transitions are equally probable. Assume now that the black hole is in
its n-th excited state. Then, the decay to the ground state through $k$
intermediate states can occur in $\frac{n!}{k!(n-k)!}$ different ways.
Summing over $k$ gives the number of possible different transitions
$N_{\mbox{\tiny transitions}}=2^n$. Thus the corresponding information
capacity is approximately
\begin{equation}
I_{\mbox{\tiny transitions}} \approx n \ln 2  \propto
       S_{\mbox{\tiny bh}}
\end{equation}
Therefore,  the mechanism proposed in \cite{bel}          could be behind
the resolution of the paradox, because enough information could be
encoded in the transitions to the ground state. Nevertheless,
if the Hawking   radiation were exactly thermal, then this mechanism
would be irrelevant because it lacks the vehicle necessary  to transmit
the information to a distant observer. However, it has recently been
shown, based on information theoretic premisses  \cite{bek},
that the fact that
the black hole absorptivity is not unity could render this radiation
the intermediary between the black hole and a distant observer.
This is so
because the radiation is not exactly thermal, i.e.  not completely
random, and there is enough thermodynamical room in the radiation
to transfer all this information.

For an observer far away from the black hole, the situation
would be quite acceptable.
The black hole appears as a quantum object emitting
Hawking radiation whose spectral lines can be used to reconstruct
all the information. The black hole is in some sense not very different
from an atom. But, contrary to the case of an atom, we can move in
closer and investigate the macroscopic black hole and its
horizon in greater detail. Then we will observe effects that  we
will experience as non-local, transmitting information from the
interior across the apparent horizon. It is important to note,
and this is precisely what we have proven quite generally
in the previous section, that this occurs {\it throughout} the history
of the evaporating black hole.  Even when it is macroscopic. There
is {\it no way}, unless we consider alternative {\bf II} above, to
delay this to the later stages of the evaporation.

The key question is: Can such processes be harmless without causing
new paradoxes? In this context we  must examine also in a more
 quantitative way
how restrictive the presence of an {\it apparent} horizon is.
Even if, as we have argued, complete reflection of information at the
macroscopic apparent horizon is impossible, it is conceivable that it
could take place at the event horizon, which might be as small as the
Planck
scale and, therefore, sensitive to quantum gravity effects. The key
question is whether this is too late, in the sense that the remaining
energy would be compatible with the information content. It is commonly
accepted that this is really too late. This is also the reason why
we have been forced to consider non-local effects. However, a more
quantitative analysis would clearly be needed to rule out this
possibility, which otherwise would make these    effects unnecessary,
or at least present only close to the event horizon and the singularity.
In fact, through redshifting, Planck scale
physics near the event horizon will be magnified tremendously in the
eyes of an observer at infinity. While the time to fall into the
 black-hole
is very short for the freely falling black-hole explorer, it would take
of the order of the whole evaporation time according for an observer at
infinity. A Planck time before the event horizon  might be well in
advance of the complete evaporation, while the black hole is still
macroscopic as viewed from the outside. A similar
suggestion has been made in \cite{ver} in the context of
two-dimensional dilaton gravity.

\section{Summary}

Our discussion points out that if we do not allow for non-unitarity,
we must {\it either}
learn to live with an infinite number of metastable or
stable black-hole remnants, {\it or} there must exist non-local
information transfer, which is at work throughout the evaporation,
even when the black hole is macroscopic.
Our conclusion is that
quantum correlations are insufficient to
solve either of these problems. In the first case, we have shown that
the information storage in correlations does  not allow us to decrease
the number of needed black-hole states. In the second case, it is
well known that EPR correlations do not allow for the kind of
information transfer that  is needed. If we say that a black hole
is like an atom with information encoded in its spectral lines,
we still need to confront the issue of locality.

\section*{Note Added}

After completion of this work we received a paper  \cite{don},
where the information paradox is discussed.

\section*{Acknowledgments:} M.S. is partially supported by World
Laboratory. M.S. is particulary indebted to Jacob Bekenstein, Maurizio
Gasperini and Don Page for enlightening conversations.

\end{document}